# Field Control of Anisotropic Spin Transport and Spin Helix Dynamics in a Modulation-Doped GaAs Quantum Well

S. Anghel,[1,*] F. Passmann,[1,*] A. Singh,[2] C. Ruppert,[1] A. V. Poshakinskiy,[3] S. A. Tarasenko,[3] J. N. Moore,[4] G. Yusa,[4] T. Mano,[5] T. Noda,[5] X. Li,[6] A. D. Bristow[1,7] and M. Betz[1]

[1] *Experimentelle Physik 2, Technische Universität Dortmund, Otto-Hahn-Straße 4a, D-44227 Dortmund, Germany*
[2] *Department of Material Science and Engineering, Massachusetts Institute of Technology, Cambridge, MA 02139, U.S.A.*
[3] *Ioffe Institute, St. Petersburg 194021, Russia*
[4] *Department of Physics, Tohoku University, Sendai 980-8578, Japan*
[5] *National Institute for Materials Science, Tsukuba, Ibaraki 305-0047, Japan*
[6] *Texas Materials Institute, University of Texas at Austin, Austin, Texas 78712, U.S.A.*
[7] *Department of Physics and Astronomy, West Virginia University, Morgantown, WV 26506-6315, U.S.A.*
E-mail address: markus.betz@tu-dortmund.de
*the authors have contributed equally to the manuscript*

Electron spin transport and dynamics are investigated in a single, high-mobility, modulation-doped, GaAs quantum well using ultrafast two-color Kerr-rotation micro-spectroscopy, supported by qualitative kinetic theory simulations of spin diffusion and transport. Evolution of the spins is governed by the Dresselhaus bulk and Rashba structural inversion asymmetries, which manifest as an effective magnetic field that can be extracted directly from the experimental coherent spin precession. A spin precession length $\lambda_{SOI}$ is defined as one complete precession in the effective magnetic field. It is observed that application of (a) an out-of-plane electric field changes the spin decay time and $\lambda_{SOI}$ through the Rashba component of the spin-orbit coupling, (b) an in-plane magnetic field allows for extraction of the Dresselhaus and Rashba parameters, and (c) an in-plane electric field markedly modifies both the $\lambda_{SOI}$ and diffusion coefficient. While simulations reproduce the main features of the experiments, the latter results exceed the corresponding simulations and extend previous studies of drift-current-dependent spin-orbit interactions.

## I. INTRODUCTION

The quantum phenomenon that is spin-orbit interaction (SOI) plays a central role in the behavior of spin transport and dynamics in semiconductors [1-3]. In non-centrosymmetric crystals, the conduction band is spin-split by Dresselhaus SOI, originating from the bulk inversion asymmetry (BIA) [4]. Dresselhaus SOI in quantum wells (QWs) encompasses contributions that are linear and cubic in the carrier momentum ***k***. Additionally, carriers in heterostructures experience Rashba SOI, which arises from the structural inversion asymmetry (SIA) of the grown layer sequence [5] and is linear in momentum ***k***. These SOI components can be tailored in a QW by choosing an appropriate confinement potential. The Rashba SOI can also be externally tuned by applying a back-gate voltage, which affects the shape of the confinement potential.

SOI leads to a ***k***-dependent effective magnetic field ***B***$_{SOI}$ that electrons feel if they propagate through the crystal, resulting in Larmor precession of the electron spins around the effective magnetic field [6,7]. ***B***$_{SOI}$ can be expressed in terms of a spin-orbit coupling parameter, which we call the spin precession length $\lambda_{SOI}$, corresponding to the distance over which the spin of a propagating electron completes one full rotation around the effective magnetic field.

SIA and BIA in [001]-oriented QWs can interfere and result in strong anisotropy of the spin splitting, particularly when the Rashba and Dresselhaus terms are of similar strength [8,9]. Under these circumstances, the Dyakonov–Perel spin relaxation mechanism is suppressed for certain spin wave modes and a persistent spin helix (PSH) is formed, as predicted [10,11] and observed in GaAs QWs [12-15]. GaAs QWs can have a long spin decay time and high electron mobilities within the regime of strong SOI making it a good choice for study of SOIs.

$\lambda_{SOI}$ incorporates both Rashba ($\alpha$) and Dresselhaus ($\beta$) parameters. Typically, the Rashba contribution is more susceptible to changes of the band structure by external fields and/or photoexcitation. Information about the temporal [16] and back-gate voltage $U_{BG}$ dependences

[13,17] of $\lambda_{SOI}$ has been obtained in recent scientific efforts addressing the PSH in QWs. In a recent study, some of the authors demonstrated that the lifetime of electron spins markedly depends on $U_{BG}$, i.e., on electron concentration [18]. However, the applied in-plane electric and magnetic field dependence of $\lambda_{SOI}$ is still incomplete, with few examples of field control of the SOI [19]. For example it is unclear how $\lambda_{SOI}$ depends on diffusion versus drift currents.

In this article, we employ a time-resolved polar magneto-optic Kerr rotation microscopy technique to perform a comprehensive investigation of the dependence of $\lambda_{SOI}$ on the in-plane electric and magnetic fields in a modulation-doped GaAs QW with high electron mobility exceeding $10^6$ cm$^2$/V s, which is an order of magnitude larger than other recent work [19]. In a regime with BIA and SIA terms slightly detuned from the PSH conditions, we directly measure the diffusive evolution of a locally excited spin ensemble into a spin helix (SH) with shorter lifetime than the PSH.

From experimental results, the dependence of $\lambda_{SOI}$ on delay time, carrier density, back-gate voltage and the in-plane crystallographic direction are determined. It is found that $\lambda_{SOI}$ substantially decreases over time while gradually approaching the SH precession length $\lambda_0$, which results from the finite laser spot size of pump and probe [20]. We also measure the dependence of $\lambda_{SOI}$ on the back-gate voltage, which simultaneously tunes the electron concentration and the out-of-plane electric field along the growth direction, revealing a mostly linear dependence. In addition, we measure the dependence of $\lambda_{SOI}$ on the carrier drift velocity $v_{dr}$ by applying an in-plane electric field, finding a pronounced influence for in-plane fields even as small as ~1 V/cm. In contrast, an applied in-plane magnetic field has minimal influence on the spin diffusion coefficient or SOI.

We support these experimental finds with a kinetic theory of spin diffusion and PSH formation, which qualitatively reproduces the majority of the features observed in the experiments. Details of the theory are presented in the Appendix.

## II. EXPERIMENTAL

The QW is grown by molecular beam epitaxy (MBE) on a highly n-doped [001] GaAs substrate. The width of the GaAs QW is $L_z$ = 15 nm, sandwiched between two Al$_{0.33}$Ga$_{0.67}$As barriers. A growth interruption for Si-delta doping provides an electron concentration of $n$ ~ 1.9x10$^{11}$ cm$^{-2}$ to form a two-dimensional electron gas (2DEG) with the estimated Fermi energy $\varepsilon_F$ ~ 7 meV in the QW, relative to the bottom of the conduction band. The 2DEG density is increased by increasing $U_{BG}$.

The MBE-grown heterostructures is processed into a Hall-bar geometry with AuGeNi Ohmic contacts. The sample is cooled to ~ 3.5 K in a helium flow cryostat (Oxford Microstat HiRes2) where the spatial fluctuations are limited to a range that is negligible in comparison to the beam size.

For the analysis of spin dynamics in the QW, we use a magneto – optical Kerr rotation (KR) setup in a confocal reflection geometry. KR is a direct measure of the dynamic magnetization in the sample and, therefore, an ideal tool to study spin phenomena in 2DEGs [21]. Briefly, the output of a mode-locked femtosecond Ti-sapphire laser of 60 MHz repetition rate is split into pump and probe pulse trains. For independent wavelength tunability, each pulse trains through its own grating-based 4f-pulse-shaper [22].

The temporal resolution of the system is 1 ps, limited by the chosen spectral width (0.7 nm). For all results presented below, the pump beam is tuned to ~1.57 eV (790 nm) with a peak power density of 4.6 MW/cm$^2$ per pulse to effectively initialize a spin polarization while the probe is set to ~1.53 eV (807 nm) with a peak power density of 2.3 MW/cm$^2$ per pulse. The polarization state of the pump pulse is modulated between $\sigma^+$ and $\sigma^-$ helicities using an electro-optical modulator (Qioptiq LM 0202P).

Both beams are collinearly focused onto the sample using a 50× objective (Mitutoyo M-Plan APO NIR) which offers a focused spot size for probe(pump) beams of ~1(3) μm. The reflected beams pass through a spectrometer, where the pump light is filtered out spectrally. The KR signal is measured in a standard balanced detection scheme. The amplitude of the signal is proportional to the rotation of the linear probe polarization caused by the magnetization associated with the pump induced spin polarization. Measurements on spin dynamics are undertaken as a function of delay time $t$ between pump and probe pulses.

In order to record the spatial spin distribution $S_z(x,y)$, a telescope scheme is implemented in the pump arm. In particular, lateral motion of one of the lenses in the pump telescope allows scanning the excitation in the $x$- and $y$- directions relative to the probe [23,24]. Here, the $x$- and $y$- directions correspond to the [1$\bar{1}$0] and [110] crystallographic directions, respectively. An electromagnet supplies a magnetic field in the Voigt geometry. For drift dependent measurements, an additional in-plane electric field is applied using the Hall-bar contacts.

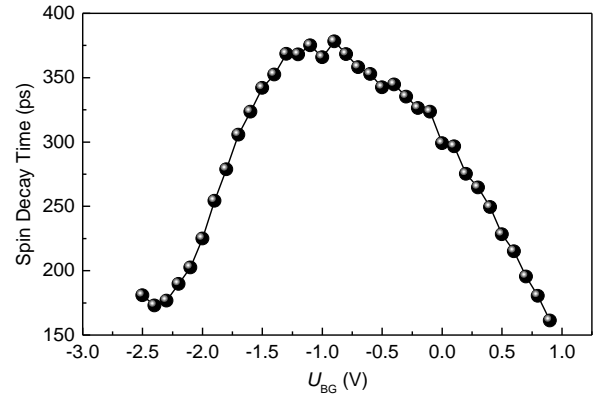

**FIG. 1** Tunability of the spin decay time in the quantum well with back-gate voltage $U_{BG}$.

## III. RESULTS AND DISCUSSION

### A. Electron-density-dependent spin lifetime

Figure 1 shows the electron spin decay time dependence on the back-gate voltage $U_{BG}$. The back-gate voltage varies the electron density $n$ within the 2DEG, in an estimated range of $10^{10} - 10^{11}$ cm$^{-2}$. In addition to which, the optical excitation also induces photocarrriers of similar and higher concentrations.

The spin decay time is maximized for an intermediate carrier density, around $-1.4$ V $< U_{BG} < -1.0$ V, a result that is comparable to recent observations in similar structures [18]. Hence, subsequent measurements will be performed with $U_{BG} = -1.4$ V, unless otherwise stated.

The non-monotonic dependence of the spin decay time with $U_{BG}$ stems from the Dyakonov-Perel spin relaxation mechanism and Coulomb screening. The rise of spin decay times in the range of $-3$ V to $-1.4$ V can be attributed to increased $n$, which slows spin relaxation due to electron-electron collisions. For increased $U_{BG}$, spin relaxation becomes faster due to the increased electron Fermi energy and suppression of electron scattering by carrier screening [18,25]. These competing mechanisms result in the observed peak of spin decay time at intermediate back-gate voltages, where interplay of the Rashba and Dresselhaus SOIs allow for the formation of a long-lived SH mode.

### B. Formation and evolution of the spin helix

The formation of a SH is conveniently visualized by tracking the spin polarization $S_z$ in space and time. This effect is extracted by keeping the probe spot fixed and scanning the pump spot along the $y$-direction for different delay times. Line scans $S_z(y,t)$ for three different values of $t$ are shown in the Fig. 2a. The first line $S_z(y,0$ ns$)$ is recorded at the temporal overlap of the pump and probe pulses. The spatial profile of the spin polarization corresponds to a convolution of pump and probe spot and is approximately Gaussian. The photoexcited electron spin ensemble expands in time due to carrier diffusion.

In Fig. 2a, $S_z(y,1$ ns$)$ and $S_z(y,1.6$ ns$)$ feature a stripe pattern consisting of an alternating sign of $S_z(y,t)$, which is caused by the spin precession around the effective magnetic field $\boldsymbol{B}_{SOI}$. A more detailed set of the $S_z(y)$ scans for a range of delay times is shown as a false-color plot in Fig. 2b. It is evident that the excited spin polarization $S$ starts with $S \parallel z$ and then oscillates as a function of $y$ as the electrons diffuse away from $y = 0$. A full oscillation of $S_z$ starts to emerge for delay times $t > 1.5$ ns. To quantify the combined effect of expanding Gaussian profiles and the SH oscillations, the experimental data $S_z(y,t)$ are fitted to the product of a Gaussian and a cosine [16,18]

$$S_z(y) = A \cdot e^{-\frac{4\ln(2)(y-y_0)^2}{w^2}} \cdot \cos\left(\frac{2\pi(y-y_1)}{\lambda_{SOI}}\right), \quad (1)$$

where $w$ and $y_0$ are the full width at half maximum (FWHM) and center positions of the Gaussian peak and $y_1$ is the spatial phase of the oscillatory function.

The width of the Gaussian envelope is expected to increase with delay time due to carrier diffusion, according to [26]

$$w^2(t) = w_0^2 + 16\ln(2)\, D_s t, \quad (2)$$

where $D_s$ is the spin diffusion coefficient, $w_0$ is the initial FWHM determined by the laser spot sizes on the sample. A linear fit of the beam waist $w^2(t)$ (data not shown) provides a direct measure of $D_s$. We obtain a value of $D_s = 57.3$ cm$^2$ s$^{-1}$ ($w_0 = 3.3$ μm) comparable to previous reports on similar QWs [27].

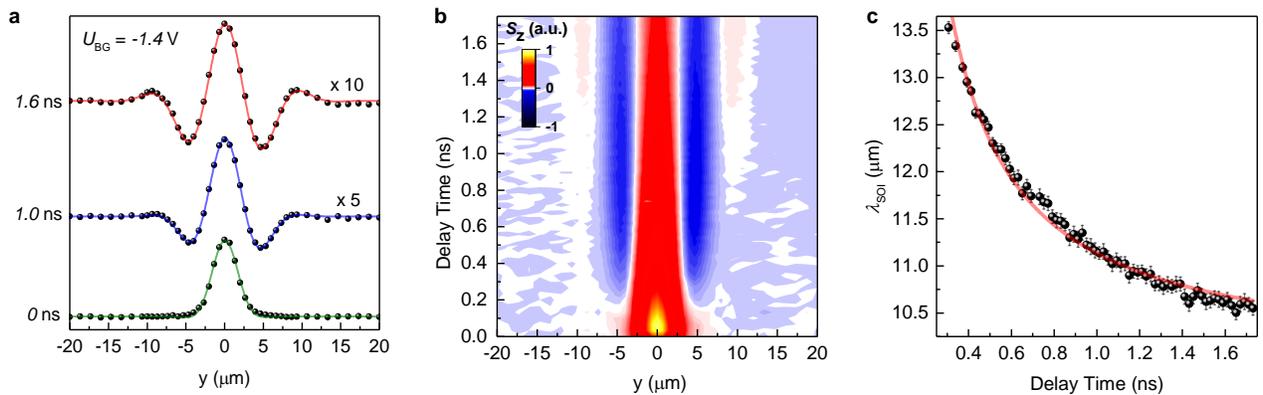

**FIG. 2 (a)** Normalized spatial spin-polarization distribution $S_z$ measured along the $y$-axis for various delay times and a constant back-gate voltage of $U_{BG} = -1.4$ V. The three data sets are normalized and vertically shifted for clarity. The solid lines represent fits according to Eq. (1). **(b)** Spatio-temporal evolution of $S_z(y,t)$ presented as false-color plot. **(c)** Temporal evolution of $\lambda_{SOI}$ extracted with Eq. (1) from (b) and a solid fit line using Eq. (3).

Figure 2c displays the transient $\lambda_{SOI}(t)$, which reveals a decrease from 13.5 µm initially, towards ~10.5 µm after a delay time of > 1 ns. This decrease is well described by the time dependence [20]

$$\lambda_{SOI}(t) = \lambda_0 \left(1 + \frac{w_0^2}{16 \cdot \ln(2) \cdot D_s \cdot t}\right) \quad (3)$$

after ~0.28 ns when the evolution of $S_z$ allows for resolution of $\lambda_{SOI}$ in Eq (1). The fit (red line) yields a precession length of the SH mode of $\lambda_0 = 9.9$ µm and a spin diffusion coefficient of $D_s = 74$ cm$^2$ s$^{-1}$; the latter is consistent with values obtained from the $w^2(t)$ dependence. Prior to addressing more detailed questions we shall discuss the physics behind this spatial spin precession.

The oscillation is dependent on the SIA and BIA contributions to SOI, arising from the geometry and material of the experiment. The lack of space inversion symmetry in zinc-blende-type structures, such as GaAs QWs, lifts the spin degeneracy of the conduction band. This phenomenon is quantified by a spin and momentum dependent contribution to the Hamiltonian that can be interpreted as an effective magnetic field $\boldsymbol{B}_{SOI}$ acting on propagating electron spins [7]. Choosing the Cartesian axes in the QW plane to be $x \parallel [1\bar{1}0]$ and $y \parallel [110]$, $\boldsymbol{B}_{SOI}$ can be written as

$$\boldsymbol{B}_{SOI} = \frac{2}{g\mu_B}\begin{pmatrix}[\alpha + \beta] \cdot k_y \\ [\beta - \alpha] \cdot k_x\end{pmatrix} \quad (4)$$

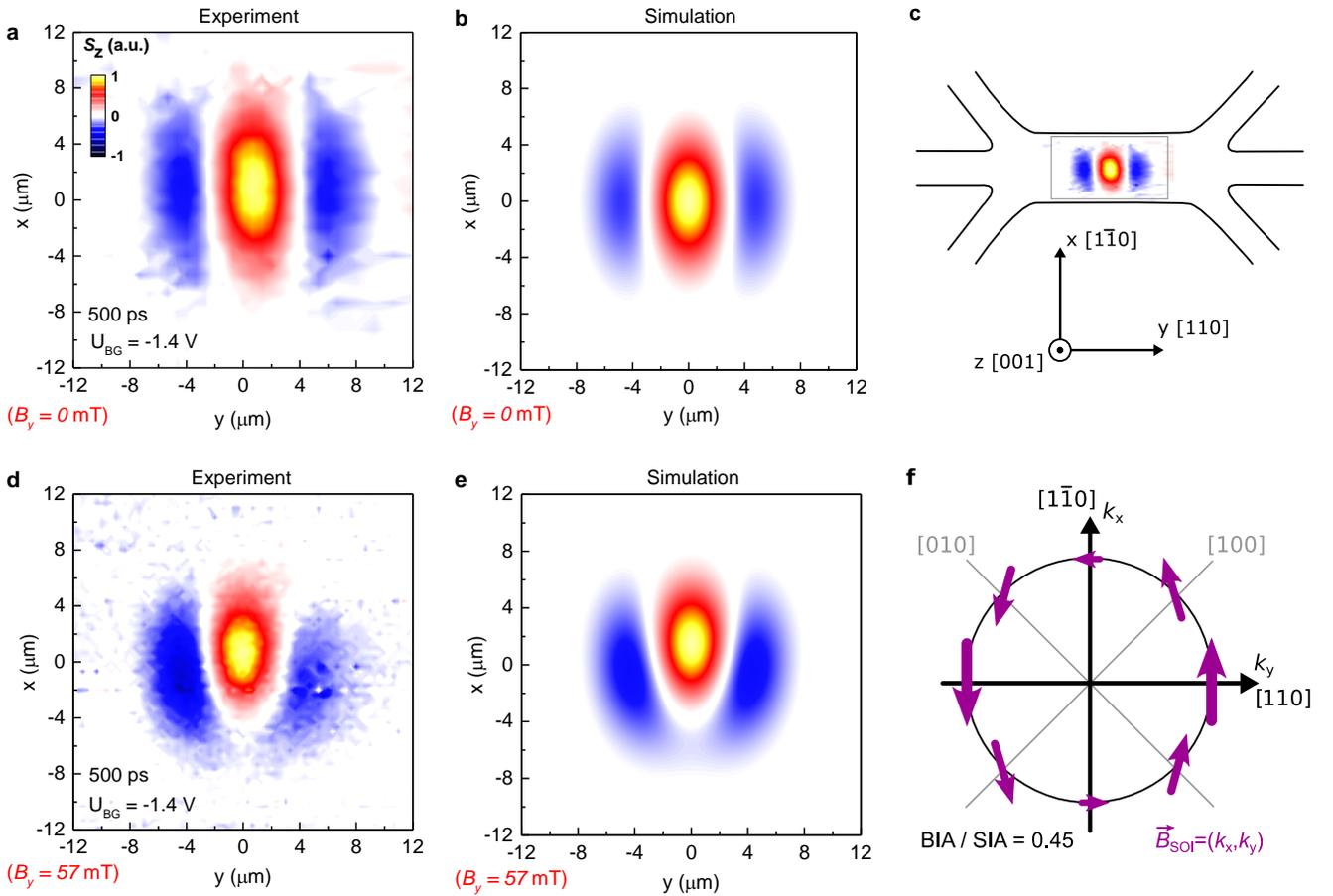

**FIG. 3** Experimentally measured and simulated two-dimensional spatial maps of the spin polarization $S_z$ due to diffusive transport. Data is obtained for the back-gate voltage of $U_{BG} = -1.4$ V and at a delay time of 0.5 ns without external magnetic field **(a)** and **(b)** and $B_y = 57$ mT **(d)** and **(e)** respectively. The pattern in the absence of the field results from the formation of a spin helix along $y$ axis, while an applied external magnetic field allows precession to take place also along the $x$-axis. The simulated maps are obtained by numerically solving Eq. (A6) for the set of parameters extracted from measurements. **(c)** Schematic visualization of the map in (a) on the sample surface (inserted false-color plot magnified for improved visualization). **(f)** Orientation and magnitude of the effective magnetic field $\boldsymbol{B}_{SOI}$ in $\boldsymbol{k}$-space on the Fermi surface, plotted for the ratio of $\alpha/\beta = 0.45$ with zero applied magnetic field, matching panels (a) and (b).

where $k_x$ and $k_y$ are the in-plane wave vectors, $g$ is the effective $g$-factor, $\mu_B$ is the Bohr magneton, and $\alpha(\beta)$ is the Rashba(Dresselhaus) parameter related to the strength of the SIA(BIA).

In the envelope function approach, $\beta = -\gamma_D[\langle k_z^2 \rangle - k_F^2/4]$, where $\gamma_D$ is the Dresselhaus coupling constant, $\langle k_z^2 \rangle = (\pi/L_z)^2 = 0.04$ nm$^{-2}$ is a lower bound for the first term and $k_F = \sqrt{2\pi n}$, such that $k_F^2/4 \sim 0.003$ nm$^{-2}$ when estimated with the zero-bias electron density. The Rashba parameter $\alpha = \gamma_R E_z$ is related to the internal electric field $E_z$ oriented in the growth direction of the QW, which can be tuned by the $U_{BG}$ through electron density and screening effects.

$\boldsymbol{B}_{SOI}$ can be examined through two-dimensional spatial maps of the spin polarization $S_z(x,y)$; see the false-color representation in Fig. 3a and its orientation with respective to the Hall-bar in Fig. 3c. The figures show the spatial spread of $S_z$ due to diffusion in the plane of the QW for $t = 500$ ps. The spatial map exhibits an oscillating pattern arising from the spin precession of diffusing carriers around $\boldsymbol{B}_{SOI}$. The stripe-like pattern in Fig. 3a clearly shows that only electrons propagating along [110], i.e., $y$-direction, undergo a significant spin precession. The underlying anisotropy of $\boldsymbol{B}_{SOI}$ in $\boldsymbol{k}$-space is visualized in Fig. 3f. This directional dependence of $\boldsymbol{B}_{SOI}$ is calculated for a ratio of the SIA and BIA contributions of $\alpha/\beta = 0.45$. This configuration results in a strong effective magnetic field felt by electrons travelling along the $y$ axis, whereas electrons traveling along the $x$ axis feel almost no $\boldsymbol{B}_{SOI}$.

When $\alpha/\beta \to 1$, at the Fermi level, the anisotropy becomes even stronger as $\boldsymbol{B}_{SOI}$ vanishes for the electrons propagating along the $x$-direction. In this balanced regime, $\alpha = \beta$ gives rise to the SU(2) spin rotation symmetry within the 2DEG, supporting a PSH and extending the spin decay time. From the absence of an oscillatory spin pattern along the $x$ axis, it is confirmed that the Rashba and Dresselhaus parameters are of the same order of magnitude for $U_{BG} = -1.4$ V.

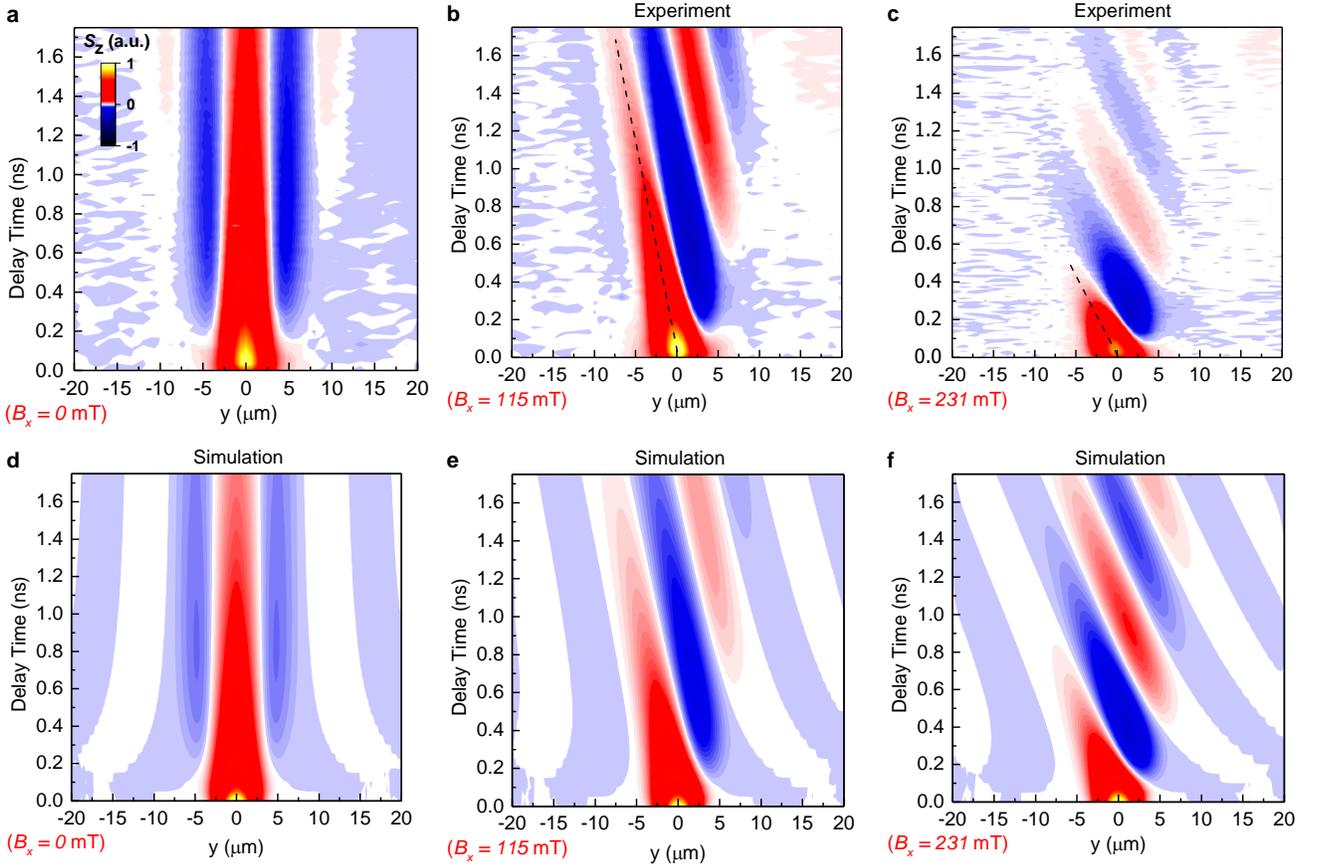

**FIG. 4** Experimental spatio-temporal maps of the evolution of $S_z(y,t)$ for $U_{BG} = -1.4$ V and $B_x$ (oriented along the [1$\bar{1}$0] direction) equal to **(a)** 0 mT, **(b)** 115 mT, and **(c)** 231 mT. The dashed lines indicate the time dependence of the center position $y_0$, which is the position of constant phase within the spin precession. Theoretical simulations, based on Eq. (A6), are displayed in **(d)**, **(e)** and **(f)** for magnetic field strengths corresponding to those in the experiments.

Parameters of the long-lived spin mode can be calculated in the framework of the kinetic theory described in the Appendix. For the diffusive transport regime and $\alpha \cdot \beta > 0$, the spin precession length is given by

$$\lambda_0^{-1} = \frac{m^*}{\pi \hbar^2} |\alpha + \beta| \sqrt{1 - \frac{1}{16}\tan^4\left(\phi - \frac{\pi}{4}\right)}, \quad (5)$$

where $m^*$ is the electrons effective mass, $\hbar$ is Plank's constant divided by $2\pi$ and the phase is $\phi = \arctan(\alpha/\beta)$. Hence, spatial mapping of the spin precession allows for a quantitative determination of $\boldsymbol{B}_{\text{SOI}}$ by the sum $|\alpha + \beta|$.

### C. Applied in-plane magnetic-field dependence

The individual $\alpha, \beta$ parameters can be extracted from the anisotropy of the two-dimensional spatial maps in the presence of an external magnetic field. Comparison of the maps with and without the external magnetic field, applied in the y-direction, shows an additional oscillation in the x-direction with the field. From the best fit of the experimental data at $B_y = 0$ (Fig. 3a) and $B_y = 57$ mT (Fig. 3d), employing the drift-diffusion equation [Eq. (A6) of the Appendix], we extract $\alpha = 1.1 \cdot 10^{-11}$ eV·cm, $\beta = 2.5 \cdot 10^{-11}$ eV·cm and $D_s = 64$ cm$^2$/s.

Figures 3b and 3e show simulations of spatial distributions of $S_z$ calculated from the extracted experimental parameters, with an effective g-factor g-factor = 0.03 and a Gaussian excitation spot with a FWHM of 3.3 μm. These simulations reproduce all the essential features of the experimental $S_z(x,y)$, demonstrating the power of the kinetic approach. It should be noted that the false-color plots for both experiment and theory have the same scale.

In the presence of an external magnetic field $\boldsymbol{B} \parallel x$, we obtain information about the spin precession in the (yz) plane. The resulting temporal evolution of $S_z(y)$ is shown in Fig. 4 for $B_x$ equal to (a) 0 mT, (b) 115 mT and (c) 231 mT. Application of $B_x$ tilts the stripe pattern observed without the field, indicating a temporal shift of a given peak position (corresponding to a fixed orientation of the spins) towards $y < 0$. Fitting the data to Eq. (1) reveals a linear dependence of the parameter $y_1$ on $t$, due to a constant velocity of the carriers in the applied field. The tilted time-evolution arises from the motion of electrons and the cancellation of $B_{\text{SOI}}(k_y)$ by $B_x$. Consequently, for an electron with the exact momentum $\hbar k_y = m^* dy_1/dt$, precession is suppressed as it propagates.

Taking into consideration Eq. (4) and the condition $B_x = B_{\text{SOI}}$, velocity in the y-direction is

$$\frac{dy_1}{dt} = \frac{\hbar g \mu_B}{2m^*(\alpha+\beta)} \cdot B_x, \quad (6)$$

As a result, the analysis of Fig. 4 gives direct access to $[\alpha + \beta]$. Specifically, we analyze the temporal derivative of the fit parameter $y_1$ for different values of $B_x$. Figure 5a shows the expected linear trend. From the slope of the linear dependence (dashed line) we find a strength of the SOI-coupling of $\alpha + \beta = 4.45 \cdot 10^{-11}$ eV cm. This value is in good agreement with the one ($\alpha + \beta = 3.85 \cdot 10^{-11}$ eV cm) obtained from Eq. (5), by using $\lambda_0 = 9.9$ μm (value previously obtained – see Eq. (3)). In addition, by comparing the decay of the signals in Fig. 4a with those in Fig. 4b and 4c, it can be seen that the spin coherence suffers from field-induced dephasing. Furthermore, the spin-diffusion coefficient is not markedly influenced by moderate values of $B_x$, see Fig. 5b. Finally, the temporal evolution of the SH density distribution $S_z(y,t)$ is reproduced by the kinetic theory simulations (Figs. 4d,e,f) for the same magnetic fields used in the experiment and using the same fit parameters as used in Fig. 3.

### D. Applied in-plane electric-field dependence

The Hall-bar structure permits a weak electric field $E_y$ to be applied in the plane of the QW. $E_y$ drives a drift current that is expected to increase the distance over which coherent spin dynamics can be measured. Figure 6a depicts spatio-temporal data for an in-plane electric field of –0.9 V/cm. Initially, the electron motion is driven by diffusion and drift currents, the latter of which drags the electrons towards the positive y-direction. There is also a slight tilt in the stripes that is a consequence of oscillation in time of the spin polarization (for fixed y), referred to as $\omega$ [19] and which is proportional to $y_1/\lambda_{SOI}$. The value of $\omega$ is expected to be $E_y$ dependent. Moreover, the tilt increases as a function of time as the periodicity of the $S_z$ increases, which is in accordance with variation in $\lambda_{SOI}(t)$ presented in Fig 2(c). Visual inspection of Figs. 6(a) and 4(a) clearly shows that $\lambda_{SOI}$ is smaller with an applied field for the same delay time.

Figure 6d shows a simulation of the spatio-temporal map with an in-plane electric field. For simplicity, this simulation uses previously determined Rashba and Dresselhaus parameters at $E_y = 0$ V/cm and $D_s = 170$ cm$^2$/s extracted at $E_y = –0.9$ V/cm. The simulation reproduces the overall +y displacement of the electron

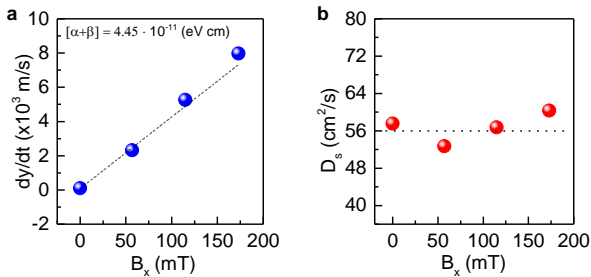

**FIG. 5 (a)** Effective velocity of the phase offset $y_1$ for several magnetic field strengths $B_x$. The dashed line is a linear fit. **(b)** Spin diffusion coefficient for different values of $B_x$. The dashed line is a guide to the eye.

ensemble due to drift, the stripes and tilt of the moving spin pattern.

Variation in $\lambda_{SOI}(t)$ with and without the applied in-plane electric field warrants direct study of $\lambda_{SOI}(E_y)$. Hence, Fig. 6b shows the dependence of $S_z(y,E_y)$ for $t = 1$ ns, when the motion is dominated only by the drift current. Electrons move towards positive $y$ values with increasing $E_y$ and produce a tilt of the striped pattern, due to an additional decrease in $\lambda_{SOI}$ with increasing $E_y$. The increased periodicity of $S_z(y,E_y)$ allows for a better fit for $w$, $\lambda_{SOI}$, $y_0$, and $y_1$ – even at small $t$. $\lambda_{SOI}(E_y)$ is extracted from the experiment by fitting $S_z(y,E_y)$ with Eq. (1) and plotted in Fig. 6c. The data support a decreasing trend of $\lambda_{SOI}$ with increasing applied field magnitude $|E_y|$ (data for $E_y > 0$ not shown).

We identify three main mechanisms that may lead to an $E_y$-dependent $\lambda_{SOI}$: (i) experimentally determined variations in $D_s$ with $E_y$ adjusts $\lambda_{SOI}$ in accordance with Eq. (3), (ii) $E_y$ alters $\alpha$ due to inadvertent components to $E_z$, and (iii) contributions to $\beta$ that depend on drift velocity (see the Supplemental Information of [19]) become significant for high mobilities. The following discussion addresses each of these mechanisms in the context of our results.

(i) Electron motion may increase spin dephasing due to an increased scattering rate. In fact, it is evident in Fig. 6b that the amplitude of $S_z(y)$ decreases with increasing drift velocity, suggesting an increased spin dephasing. Within the Dyakonov-Perel mechanism, the decrease in spin decay time for fixed SOI parameters, suggests an increase of the spin diffusion coefficient $D_s$ [28]. Such an increase of $D_s$ resulting from the in-plane electric field is indeed observed in the experiment; see Fig. 6f. The in-plane electric field dependence of $D_s$ is fit by a second order polynomial function, matching the increase of $D_s$ by a factor of three at $E_y = -1$ V/cm. According to Eq. (3) the threefold change of $D_s$ would only lead to a 3% decrease of $\lambda_{SOI}$, which is insufficient to explain the results seen in Fig. 6c.

For comparison, $D_s$ remains approximately two orders of magnitude lower than the "charge" diffusion coefficient

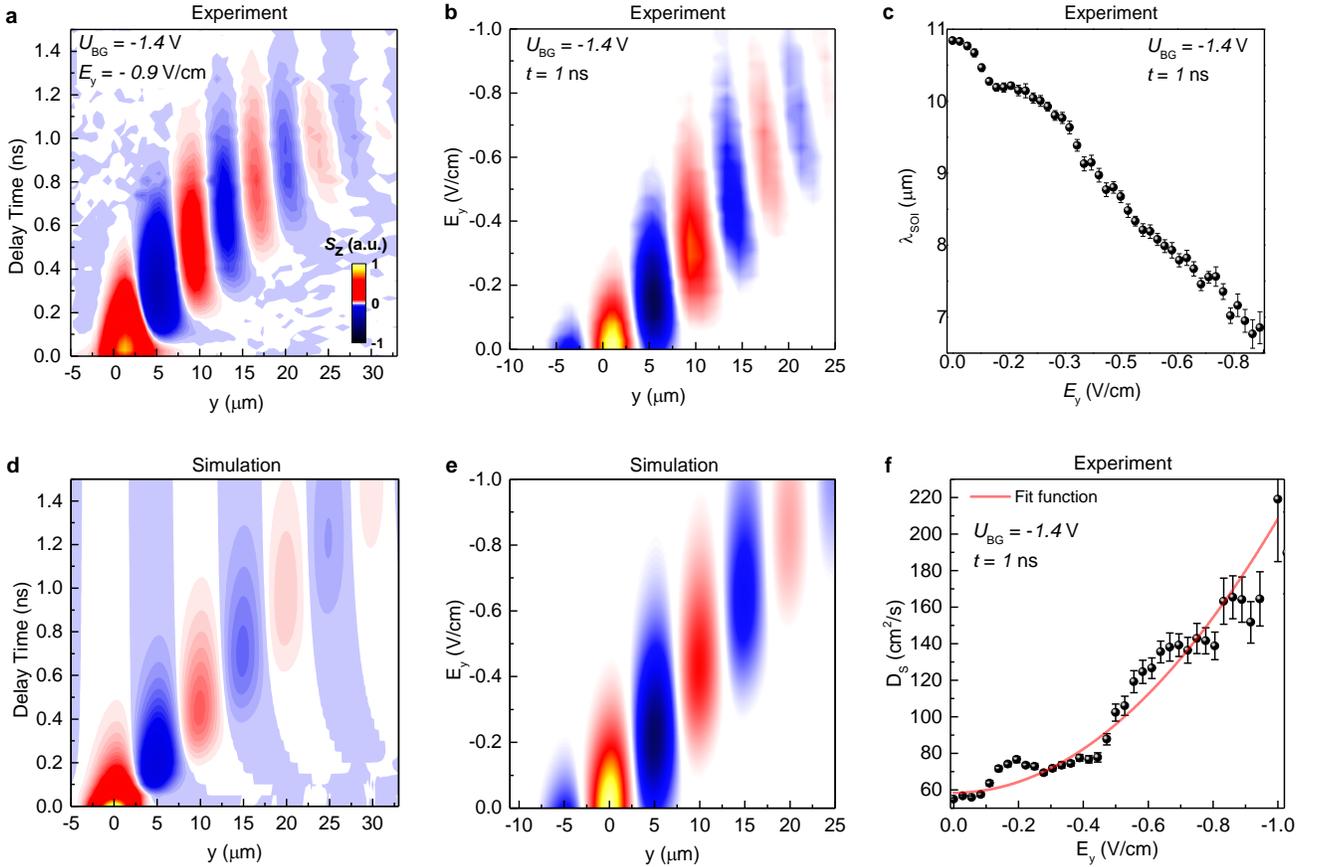

**Fig. 6** In-plane electric field dependence of the electron spin evolution: **(a)** Experimental and **(d)** simulated false-color plots of $S_z(y,t)$ for $E_y = -0.9$ V/cm. **(b)** Experimental and **(e)** simulated plots of $S_z(y,E_y)$ for $t = 1$ ns. Experimental results recorded with $U_{BG} = -1.4$ V and simulations determined from Eq. (A6). **(c)** Electric field dependence of $\lambda_{SOI}$ extracted by fitting the data of (b) using Eq. (1). **(f)** Dependence of the spin diffusion coefficient $D_s$ on $E_y$, fit by a second order polynomial function (red curve).

$D_e \sim 10^4$ cm$^2$/s, indicating the role of electron-electron scattering [29,30]. $D_e$ is estimated from the mobility $\mu \sim 1.5 \cdot 10^6$ cm$^2$/(Vs) and the Fermi energy $\varepsilon_F = 7$ meV.

(ii) The sample geometry could impart a component of the in-plane electric field into the growth direction, which potentially alters $\lambda_{SOI}$ through the Rashba component that depends on $E_z$. To verify the likelihood of this idea, we examine the dependence of $\lambda_{SOI}$ on the back-gate voltage. Figure 7a shows $S_z(y, U_{BG})$ at $t = 500$ ps and $E_y = -1$ V/cm. In this measurement, $E_y$ is applied to ensure electron transport and access $\mu(U_{BG})$.

In the range $-2.9$ V $< U_{BG} < -1.9$ V, $y_0$ remains close to $y = 0$, which is attributed to localization of the remaining electron in the QW that do not return to the doping layer in the presence of strong $E_z$. This is seen spectroscopically as a recovery of neutral exciton features in photoluminescence [18]. For $U_{BG} < -1.9$ V, $y_0$ becomes positive as the electron concentration increases and delocalization occurs. The change in $y_0$ is exhibited as the striped pattern in $S_z(y, U_{BG})$. This picture is corroborated by the dependence of $\mu(U_{BG})$; see Fig 7b. Mobility is determined using $\mu = v_{dr}/E_y$, where the drift velocity $v_{dr} = y_0/t$ is experimentally extracted from $S_z(y)$ by tracking $y_0$ with Eq. (1). It is observed that $\mu$ is small for $U_{BG} < -1.9$ V as a result of localization, above this voltage $\mu$ grows $>50\times$ as delocalization occurs.

Figure 7c shows the dependence of $\lambda_{SOI}$ on $U_{BG}$, determined by fitting $S_z(y, U_{BG})$ with Eq. (1). The magnitude of $\lambda_{SOI}$ decreases almost linearly as $U_{BG}$ is tuned from 0 V to –2 V. This behavior can be attributed to the Rashba contribution of $\lambda_{SOI}$, which is linearly dependent on the applied $E_z$, as described in relation to Eq. (4). Simultaneously, $U_{BG}$ changes the electron density $n$, which might result in an additional contribution to $\alpha$ due to Coulomb screening. Consequently, the nonlinear behavior of $\lambda_{SOI}(U_{BG})$ may result from the interplay of these inseparable contributions.

Regardless of the existence of any component of $E_y$ on $U_{BG}$, the observed change of $\lambda_{SOI}$ on $U_{BG}$ is rather small compared to its direct dependence on $E_y$. For a change in $E_y = 0.8$ V/cm, which corresponds to an in-plane bias of 1.3 V, there is a 4.5 μm change in $\lambda_{SOI}(E_y)$. By contrast $\lambda_{SOI}(U_{BG})$ shows only a 3 μm change for a 3 V change in $U_{BG}$. Hence, variation of the spin precession length due to the in-plane electric field is a more significant effect than that caused by $U_{BG}$. This confirms that $E_y$ indeed directly modifies the spin precession length.

(iii) $\boldsymbol{B}_{SOI}$ can include terms that depend on the electronic drift and diffusion currents via the Dresselhaus SOI [19]. Consequently, $\boldsymbol{B}_{SOI}$ and hence $\lambda_{SOI}$ can be modified by sufficiently large drift velocity. $v_{dr} = \mu E_y \sim 20$ km/s at $|E_y| = 1$ V/cm, which is only one order of magnitude smaller than the Fermi velocity $v_F = \hbar k_F/m^* = 297$ km/s, determined from $\beta_3 = \hbar^2\omega/(2m^* v_{dr}) = 8.2\times10^{-12}$ eVcm giving $n = 4.7\times10^{11}$ cm$^{-2}$, and an order of magnitude larger than the initial diffusion velocity $v_{Di} \sim 4$ km/s, determined from differentiation of the evolution of the half-width half maximum $w(t)/2$ of Fig 2b.

As pointed out in the Supplementary Information of Ref. [19], higher-order harmonic contributions to $\omega$ are small, estimated to be on the order of 4% due to the ratio $(k_{dr}/k_F)^2$. While these contributions are neglected, it is possible that they play a role when the drift velocity is sufficiently high or the electron sheet density is small. However, the drift-velocity-dependent contributions to $\boldsymbol{B}_{SOI}$ are insufficiently large to account for the observed $\lambda_{SOI}(E_y)$.

## IV. CONCLUSIONS

In summary, we have studied the anisotropic spin transport and spin helix formation in a modulation-doped GaAs-based quantum well. Spatio-temporal analysis of the spin polarization as a function of back-gate voltage and in-

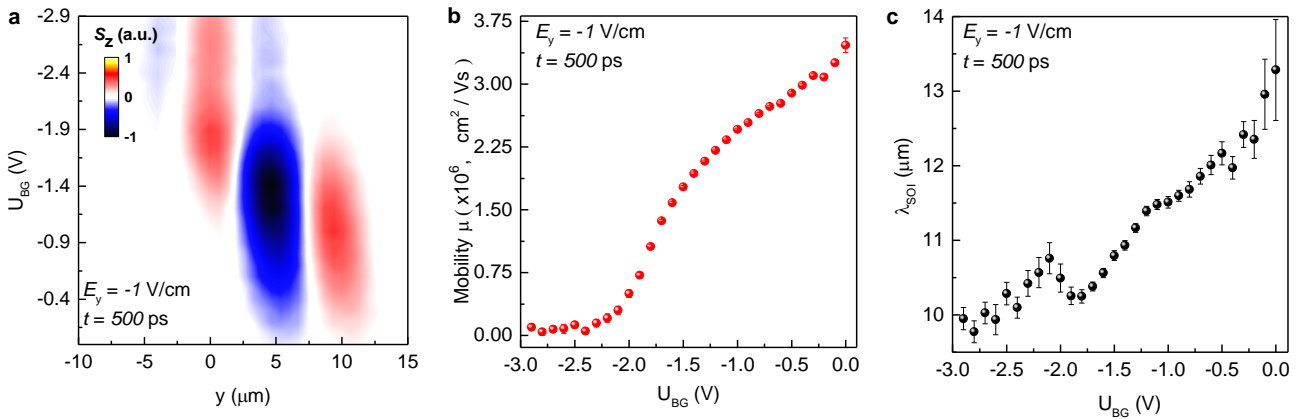

**FIG. 7** (a) Spatial evolution of $S_z(y, U_{BG})$ for in-plane electrical field $E_y = -1.0$ V/cm and delay time of 500 ps. Extracted values of back-gate voltage-dependent (b) electron mobility $\mu$ and (c) spin precession length $\lambda_{SOI}$.

plane electric field reveals a tunability of the spin precession length $\lambda_{\text{SOI}}$ with both in-plane and out-of-plane electric fields. The decay of $\lambda_{\text{SOI}}$ with delay time is correlated with the survival of the long-lived spin helix mode.

Tuning the back-gate voltage induces changes to out-of-plane electric field that results in a mostly linear dependence of $\lambda_{\text{SOI}}$. In addition, a pronounced decrease of $\lambda_{\text{SOI}}$ occurs when applying an in-plane electric field. This later result warrants further study because the change of $\lambda_{\text{SOI}}$ cannot be attributed to direct or indirect modification of the spin-orbit coupling by an altered diffusion coefficient. The observed three-fold increase in the spin diffusion coefficient with the applied in-plane electric field, is also insufficient to modify the observed change in $\lambda_{\text{SOI}}$. In contrast, application of a moderate external magnetic field does not significantly influence the spin-orbit interactions or the spin-diffusion coefficient.

## ACKNOWLEDGMENTS

We acknowledge funding from the Deutsche Forschungsgemeinschaft (DFG) in the framework of the ICRC – TRR 160, project B3. This work was also supported by the Asahi Glass Foundation and a Grant-in-Aid for Scientific Research (No. 17H01037) from the Ministry of Education, Culture, Sports, Science, and Technology (MEXT), Japan. Theoretical research was supported by the Government of the Russian Federation (project #14.W03.3.0011) and also RFBR-DFG (project 15-52-12012). The work performed by UT-Austin is supported by NFS DMR-1306878.

## APPENDIX

The experimentally observed spin diffusion and spin helix formation can be described in the framework of the kinetic equation for the spin distribution function in the wave vector and real spaces $s(k,r,t)$ that has the form

$$\frac{\partial s}{\partial t} + \frac{e}{\hbar}\left(E \cdot \frac{\partial}{\partial k}\right)s + \left(v_k \cdot \frac{\partial}{\partial r}\right)s = (\Omega_k + \Omega_L) \times s + \text{St}[s] \quad (A1)$$

where $e$ is the electron charge, $v_k = \hbar k/m^*$ is the electron velocity, $\Omega_k(\Omega_L)$ is the spin precession frequency around the effective magnetic field (external magnetic field) and $\text{St}[s_k]$ is the collision integral. The collision integral describes the scattering of electrons by QW imperfections as well as electron-electron scattering. In the collision-dominated regime, when $\Omega_k \tau \ll 1$, where $\tau$ is the scattering time, the spin distribution function is given by the sum of $k$-isotropic and $k$-anisotropic terms, $s(k,r,t) = \bar{s}(\varepsilon_k, r, t) + \delta s(k, r, t)$ with $\delta s \ll \bar{s}$. Here, the bar denotes averaging over the directions of the wave vector $k$ and $\varepsilon_k = \hbar^2 k^2/(2m^*)$ is the electron energy. We assume the energy relaxation is faster than the spin relaxation so that the electron system is thermalized and the $k$-isotropic part of the spin distribution function has the form

$$\bar{s}(\varepsilon, r, t) \propto S(r,t)\frac{d}{d\varepsilon}f_{\text{FD}}(\varepsilon), \quad (A2)$$

where $S(r) = \sum_k \bar{s}(\varepsilon_k, r)$ is the local spin density and $f_{\text{FD}}(\varepsilon)$ is the Fermi-Dirac function. Summing up Eq. (A1) over $k$ and neglecting spin relaxation due to the scattering we obtain the equation for the spin density

$$\frac{\partial S}{\partial t} = \Omega_L \times S + \sum_k \left[\Omega_k \times \delta s - \left(v_k \cdot \frac{\partial}{\partial r}\right)\delta s\right] \quad (A3)$$

The anisotropic part of the spin distribution function is found from the equation

$$\text{St}[\delta s] = \left(v_k \cdot \frac{\partial}{\partial r}\right)\bar{s} + \frac{e}{\hbar}\left(E \cdot \frac{\partial}{\partial k}\right)\bar{s} - \Omega_k \times \bar{s} \quad (A4)$$

which, in the $\tau$-approximation, yields

$$\delta s = \tau_p^* \Omega_k \times \bar{s} - \tau_p^*\left(v_k \cdot \frac{\partial}{\partial r}\right)\bar{s} - \tau_p e(E \cdot v_k)\frac{\partial \bar{s}}{\partial \varepsilon}. \quad (A5)$$

Here, $\tau_p$ is the momentum relaxation time, which determines the electron mobility and is governed by electron scattering from QW imperfections, and $\tau_p^*$ is the scattering time, which determines spin diffusion and is limited by both electron scattering from QW imperfections and electron-electron scattering [29]. Combining Eqs. (A3) and (A5) we obtain the drift-diffusion equation for the spin density

$$\frac{\partial S}{\partial t} + \left(v_{\text{dr}} \cdot \frac{\partial}{\partial r}\right)S$$
$$= D_s \frac{\partial^2 S}{\partial r^2} - \Gamma S - \left(\Lambda \frac{\partial}{\partial r}\right) \times S + (\Omega_L + \Omega_{\text{dr}}) \times S. \quad (A6)$$

Here, $v_{\text{dr}} = \langle \partial(\varepsilon \tau_p)/\partial \varepsilon \rangle eE/m^*$ is the drift velocity, $D_s = \langle v^2 \tau_p^* \rangle/2$ is the spin diffusion coefficient,

$$\Gamma_{\alpha\beta} = \langle \overline{(\Omega_k^2 \delta_{\alpha\beta} - \Omega_{k,\alpha}\Omega_{k,\beta})}\tau_p^* \rangle \quad (A7)$$

is the Dyakonov-Perel spin-relaxation-rate tensor;

$$\Lambda_{\alpha\beta} = 2\langle \overline{\Omega_{k,\alpha} v_{k,\beta}} \tau_p^* \rangle \quad (A8)$$

is the tensor describing the spin precession during diffusion;

$$\Omega_{\text{dr}} = \langle \frac{d}{d\varepsilon}\tau_p \overline{(eE \cdot v_k)\Omega_k} \rangle \quad (A9)$$

is the spin precession frequency during drift, and the angle brackets denote the energy averaging,

$$\langle A \rangle = \frac{\int A \frac{df_{\text{FD}}}{d\varepsilon} d\varepsilon}{\int \frac{df_{\text{FD}}}{d\varepsilon} d\varepsilon}. \qquad (A10)$$

By solving Eq. (A6) with the given spin-orbit coupling parameters one can calculate the temporal evolution and spatial distribution of the electron spin density, also in the presence of external electric and magnetic fields. The results of such numerical solutions for the parameters extracted from experimental data are plotted in Figs. 3, 4 and 6 and demonstrate good agreement with the experiments.

For [001]-oriented QWs, the effective spin-orbit field lies in the QW plane and is given by Eq. (4). The presence of both Rashba and Dresselhaus SOI terms leads to an anisotropy of the spin splitting and the spin helix formation. At long enough delay times, the spatial distribution of spin density is determined by spin density waves with the longest lifetime. In particular, in the absence of external electric and magnetic fields, the longest lifetime (for $\alpha \cdot \beta > 0$) is achieved for the spin density wave along $y$ with the wave vector

$$q_0 = \frac{2m^*}{\hbar^2}|\alpha + \beta|\sqrt{1 - \frac{1}{16}\tan^4\left(\phi - \frac{\pi}{4}\right)} \qquad (A11)$$

where $\phi = \arctan|\alpha/\beta|$ [3]. The corresponding length of spin precession is given by $\lambda_0 = 2\pi/q_0$.

The in-plane magnetic field modifies the spin density distribution. In a strong enough magnetic field, the longest lifetime is achieved for the mode with the wave vector $\boldsymbol{q}$, for which the direction of the effective spin-orbit magnetic field coincides with the direction of the external magnetic field $\boldsymbol{\Lambda q} \parallel \boldsymbol{\Omega}_L$. For the external magnetic field pointing along the $x$ axis, the long-lived spin density wave propagates along the $y$ axis and has the wave vector

$$q_0(\boldsymbol{\Omega}_L \parallel x) = \frac{2m^*}{\hbar}|\alpha + \beta| \qquad (A12)$$

while for the magnetic field oriented along the $y$ axis the wave propagates along the $x$ axis and has the wave vector

$$q_0(\boldsymbol{\Omega}_L \parallel y) = \frac{2m^*}{\hbar}|\alpha - \beta|. \qquad (A13)$$

Thus, from the periods of spatial spin oscillations at zero magnetic field and magnetic field pointed along $y$ axis we can determine both Rashba and Dresselhaus constants independently.